\documentclass[aps,prd,secnumarabic,amssymb, amsmath,nobibnotes,nofootinbib,11pt]{revtex4}
\usepackage{graphicx}
\usepackage{amsmath}
\usepackage{amssymb}

\begin{document}
\title{%
$\kappa$-Poincar\'e as a symmetry of flat quantum spacetime}
\author{Jerzy Kowalski-Glikman }
\email{jerzy.kowalski-glikman@ift.uni.wroc.pl}\affiliation{Institute
for Theoretical Physics, University of Wroc\l{}aw, Pl.\ Maksa Borna
9, Pl--50-204 Wroc\l{}aw, Poland, EU}
\begin{abstract}
In this short note, based on the talk given at the 3rd Conference of the Polish Society on Relativity, I present the basic points of our recent paper ``Symmetries of quantum spacetime in three dimensions'' \cite{Cianfrani:2016ogm}, stressing their physical meaning, and avoiding technical details.

\end{abstract}
\maketitle

$\kappa$-Poincar\'e algebra was first derived more than 25 years ago in the seminal papers by Lukierski, Nowicki, Ruegg, and Tolstoy \cite{Lukierski:1991pn}, \cite{Lukierski:1991ff}, and took its final form in the work of Majid and Ruegg \cite{Majid:1994cy} a few years later. It forms a (nearly unique) quantum-deformed counterpart of the Poincar\'e algebra, the algebra of relativistic symmetries of the classical Minkowski spacetime. The deformation parameter $\kappa$ of $\kappa$-Poincar\'e algebra has the dimension of mass, and it was clear from the very start that this deformed algebra, if indeed realized in nature, must somehow originate from quantum gravity, and that the deformation parameter $\kappa$ should be identified with Planck mass. More specifically, it was claimed some time ago \cite{AmelinoCamelia:2011bm} that $\kappa$-Poincar\'e should be related to the so-called Relative Locality limit of quantum gravity, in which the relevant length scale is much larger than the Planck length scale, while the energies are still Planckian. $\kappa$-Poincar\'e algebra has been also used to formulate (and investigate properties of) some test theories to be used within the quantum gravity phenomenology research programme \cite{AmelinoCamelia:2008qg}.

In spite of this expected and exploited relations with quantum gravity, surprisingly enough, $\kappa$-Poincar\'e algebra was never derived in the quantum gravity framework. It was constructed as a self-consistent (and no doubt beautiful) algebraic structure, but it relation to the symmetries of quantum spacetime were for a long time not understood. In the recent paper \cite{Cianfrani:2016ogm}  we proved, using the non-perturbative loop quantum gravity techniques that indeed $\kappa$-Poincar\'e algebra {\em is the algebra of symmetries of flat quantum spacetime} of three dimensional Euclidean quantum gravity. Let me describe how this result comes about (all the technical details can be found in \cite{Cianfrani:2016ogm}.)

Let me start with explaining the meaning of the phrase {\em symmetries flat quantum spacetime}. In classical general relativity any spacetime can be identified with some particular configuration of the gravitational field. In particular the flat Minkowski spacetime is the configuration characterized by Minkowski metric $\eta_{\mu\nu} = (1,-1,-1, \ldots)$ and the maximal sets of its Killing vectors, corresponding to the ordinary translations and Lorentz transformations, satisfying the standard Poincar\'e algebra.

This algebra can be also derived from dynamics of the gravitational field. Since gravity is a gauge theory with reparametrization invariance i.e., with vanishing Hamiltonian, its dynamics is completely characterized by the system of constraints and the algebra that the constraints satisfy. In the case of Einstein gravity (in 3D) we have to do with three constraints: two diffeomorphism constraints ${\cal D}[\vec{f}]$,  and the Hamiltonian constraint ${\cal H}[g]$, with $\vec{f}$ and $g$ being the appropriate smearing functions. These constraints satisfy the Poisson bracket algebra
\begin{align}
&\big\{{\cal D}[f_1],\, {\cal D}[f_2]\big\}={\cal D}\big[[f_1,\, f_2]\big]\\
&\big\{{\cal D}[f],\, {\cal H}[g]\big\}={\cal H}[f^a\partial_ag]\\
&\big\{{\cal H}[g_1],\, {\cal H}[g_2]\big\}={\cal D}[f(g_1,g_2)]
\end{align}
where
\begin{equation}
[f_1,\, f_2]=f_1^a\,\partial_a\vec{f}_2-f_2^a\,\partial_a\vec{f}_1\qquad
f^a(g_1,g_2)=h^{ab}(g_1\,\partial_bg_2-g_2\,\partial_bg_1)\,,
\end{equation}
and $h^{ab}$ is the inverse spacial metric. It can be checked that if the smearing functions are the Killing vectors of the flat Minkowski space and the metric $h^{ab}$ is the Minkowski space metric the algebra above becomes the Poincar\'e algebra. In fact, the analogous statement holds for other maximally symmetric spaces, de Sitter and Anti de Sitter.

Now it is clear how the quantum analogue of the flat space and its symmetries can be defined. One has to replace the constraints with the appropriate constraints operators, compute their commutators, corresponding to the Poisson bracket algebra (1)--(3) with the same smearing functions being the Killing vectors and then interpret the result.

This can be  done rigorously in the context of Euclidean gravity in 3D. However the procedure is not completely straightforward. In 3D it is convenient to formulate gravity as Chern-Simons gauge theory of the de Sitter, Anti de Sitter, or Poincar\'e gauge group, depending on the value of cosmological constant. The constraints structure in Chern-Simons theory is closely related, but in some aspects drastically different from the constraints in the metric formulation of gravity, and for an arbitrary configuration of the gravitational field it is very hard to construct the quantum constraint operators of the metric theory from the Chern-Simons ones. Fortunately, as shown in \cite{Cianfrani:2016ogm}, when the gravitational field is in the maximally symmetric state this problem disappears and it is actually possible to find the one-to-one correspondence between the constraint operators in these two formulations of the theory.

For technical reasons, we are able to compute the algebra of constraints in one particular case, which is the case of Euclidean quantum gravity in 3D, with positive cosmological constant. In this case the algebra of classical constraints is $SO(4)$, which decomposes into two commuting $SU(2)$ sectors and it turns out that also the algebra of quantum constraints is a direct sum of two subalgebras. We will concentrate on one of them for a while and we will return to the whole algebra later.

The algebra of the quantum constraint operators belonging classically to the $SU(2)$ sector was computed some time ago in \cite{Pranzetti:2014xva}. It turns out that this algebra is anomalous in general, which did not come as a big surprise, since it was expected that the algebra of constraints of quantum gravity might have this property. Remarkably however, it turns out that the quantum algebra is anomaly-free if the trace of the holonomy satisfies the condition $\mbox{tr}[W_p]=-(A^2 + A^{-2})$, where $A=\exp{i\hbar\sqrt{\Lambda}/4\kappa}$,  $\Lambda$ is the cosmological constant and $\kappa$, the inverse of the Newton's constant is the Planck mass (in 3D). When this condition is met, the algebra of constraints becomes identical to the algebraic sector of the Hopf $SU_q(2)$ algebra.

However the fact that the algebra structures of two algebras are identical does not mean that the two algebras are identical as Hopf algegras. In fact, nontrivial Hopf algebras are characterized by non-trivial R-matrices, telling for example how the crossing operator acts. In our case the operators of constraints are closely associated with the holonomy operators that create a Wilson line. When two such operators act on the vacuum one after another we have to do with two lines, crossing each other at some point. By analysing the action of these operators it is possible to derive the form of the R-matrix and to check that the so obtained R-matrix is indeed the one of the Hopf algebra $SU_q(2)$ with $q=\exp{i\hbar\sqrt{\Lambda}/2\kappa}$. Using the same reasoning in the case of the remaining constraints belonging, classically, to another $SU(2)$ sector, we find the Hopf algebra $SU_{q^{-1}}(2)$. Thus the algebra of quantum constraints of the Euclidean de Sitter space in 3D quantum gravity becomes $SU_q(2)\oplus SU_{q^{-1}}(2)$.

Having obtained this result we are facing two questions. First, the original group of classical spacetime symmetries for Euclidean de Sitter space is $SO(4)$ that decomposes into the direct sum $SU(2)\oplus SU(2)$. The question arises as to if also in the quantum case the direct sum $SU_q(2)\oplus SU_{q^{-1}}(2)$ can be combined into a single Hopf algebra? Second, what about the flat space, vanishing cosmological constant limit? Does it exist? Is it non-trivial?

All these question has been answered in the affirmative, although the explicit constructions are quite delicate. The direct sum $SU_q(2)\oplus SU_{q^{-1}}(2)$ indeed combines into a particular real form of the Hopf algebra $SO_q(4, \mathbb{C})$. And also the flat limit exists, and is non-trivial: it is given by Euclidean, 3D, $\kappa$-Poincar\'e algebra. Thus indeed, $\kappa$-Poincar\'e is the {\em deformed} symmetry of flat quantum spacetime of 3D gravity. It should be stressed that this result is made possible by neat cancellations resulting from the actual decomposition structure, in which the deformation parameter of one sector is $q$ and another, its inverse $q^{-1}$.

Let me finish this short note with some comments regarding the directions of the follow-up research. Certainly it would be of great interest to extend this result to other cases of different signs of the cosmological constants and/or spacetime metric signature. Although the algebraic part of this procedure is fully under control, it is not clear how to extend the basic computation reported in \cite{Pranzetti:2014xva}. Even more challenging will be to extend the results obtained to the case of quantum gravity in physical four spacetime dimensions. However, our result strongly suggests that also in these cases the effective flat quantum spacetime algebra of symmetries must be deformed and it is likely that it has the form of the appropriate $\kappa$-Poincar\'e algebra.

\section*{Acknowledgements}
This work was supported  by funds provided by the National Science Center, projects number 2011/02/A/ST2/00294 and
2014/13/B/ST2/04043.

\end{document}